\documentclass[preprint,amsmath,amssymb,showkeys]{revtex4-1}
\usepackage{graphicx}
\usepackage{xcolor}
\usepackage{enumerate}
\usepackage{longtable}

\begin{document}

\title{A universal approach for drainage basins}

\author{Erneson A. Oliveira$^{1,2,3}\footnote{Correspondence to:
erneson@eaoliveira.com}$, Rilder S. Pires$^{3}$, Rubens S. Oliveira$^{3}$, Vasco
Furtado$^1$, Hans J. Herrmann$^{3,4,5}$, Jos\'e S. Andrade Jr.$^{3}$}

\affiliation{$^1$ Programa de P\'os Gradua\c c\~ao em Inform\'atica Aplicada, Universidade de Fortaleza, 60811-905 Fortaleza, Cear\'a, Brasil.\\
$^2$ Mestrado Profissional em Ci\^encias da Cidade, Universidade de Fortaleza, 60811-905 Fortaleza, Cear\'a, Brasil.\\
$^3$ Departamento de F\'isica, Universidade Federal do Cear\'a, Campus do Pici, 60451-970 Fortaleza, Cear\'a, Brasil.\\
$^4$ PMMH, ESPCI, 7 quai St Bernard, 75005 Paris, France.\\
$^5$ ETH Z\"urich, Computational Physics for Engineering Materials, Institute for Building Materials, Wolfgang-Pauli-Strasse 27, HIT, CH-8093 Z\"urich, Switzerland.}

\date{\today}

\begin{abstract}
Drainage basins are essential to Geohydrology and Biodiversity. Defining those
regions in a simple, robust and efficient way is a constant challenge in Earth
Science. Here, we introduce a model to delineate multiple drainage basins
through an extension of the Invasion Percolation-Based Algorithm (IPBA). In
order to prove the potential of our approach, we apply it to real and artificial
datasets. We observe that the perimeter and area distributions of basins and
anti-basins display long tails extending over several orders of magnitude and
following approximately power-law behaviors. Moreover, the exponents of these
power laws depend on spatial correlations and are invariant under the landscape
orientation, not only for terrestrial, but lunar and martian landscapes. The
terrestrial and martian results are statistically identical, which suggests that
a hypothetical martian river would present similarity to the terrestrial rivers.
Finally, we propose a theoretical value for the Hack's exponent based on the
fractal dimension of watersheds, $\gamma=D/2$. We measure $\gamma=0.54 \pm 0.01$
for Earth, which is close to our estimation of $\gamma \approx 0.55$. Our study
suggests that Hack's law can have its origin purely in the maximum and minimum
lines of the landscapes.
\end{abstract}
\maketitle

\section{Introduction}

Drainage basins play a fundamental role in the hydrologic cycle, which make them
essential to diversity and maintenance of Life on Earth \cite{fetter2014}. The
longstanding problem of characterising drainage basins has drawn much attention
due to its importance in a variety of environmental issues, such as water
management \cite{vorosmarty1998, knecht2012, brooks2012}, landslide and flood
prevention \cite{dhakal2004, pradhan2006, lazzari2006, lee2006, yang2007}, and
aquatic dead zones \cite{diaz2008, breitburg2018, goudie2018}. In this context,
{\it drainage basins}, or simply {\it basins}, are all land areas sloping toward
a single outlet, {\it e.g.} a river mouth or points of higher infiltration or
evaporation rates. They are outlined by abstract boundary lines, called {\it
topographic divides} or {\it watersheds}. The concept of watersheds appears in
many other seemingly unrelated areas like percolation theory \cite{araujo2014,
saberi2015}, image segmentation and medicine \cite{vincent1991, grau2004,
ng2006, cousty2010}, and even international borders \cite{un1902, edaa2009}.

Watersheds are recognized as fractals in several cases \cite{breyer1992,
fehr2009, fehr2011a, fehr2011b}, exhibiting self-similarity and a well defined
{\it fractal dimension}. There are several objects that share the same fractal
dimension of watersheds on uncorrelated random substrates ({\it viz.} $D\approx
1.22$), such as optimal paths under strong disorder \cite{cieplak1994,
cieplak1996, porto1997, porto1999}, minimum spanning trees on random networks
\cite{dobrin2001}, backbones of the optimal path crack \cite{andrade2009,
oliveira2011, andrade2011}, and bridge bonds on ranked surfaces
\cite{shrenk2012}. All these loopless paths belong to the same universality
class as watersheds. Furthermore, it is also known that watersheds are
Schramm-Loewner evolution curves \cite{daryaei2012} and that it is possible to
define hydrological watersheds \cite{burger2018}, where the infiltration process
in the soil is taken into account.

The availability of digital elevation models (DEMs) allowed the development of
modern techniques to automatically delineate watersheds. Nowadays, these methods
are used in the standard Geographic Information System (GIS) software and are
well established as a fundamental tool in Geoscience \cite{schwanghart2014}. The
basic idea behind these methods is the calculation of the flow directions, which
can be defined in several ways depending on the assumptions on the flow and
number of neighbouring cells in the grid \cite{ocallaghan1984, freeman1991,
quinn1991, costa-cabral1994, tarboton1997, peckham2007, gruber2009, shelef2013}.
Simultaneously to the development of these methods, the concept of watersheds
were also introduced in the context of contour delineation \cite{beucher1979}
and became a widely used tool for image processing. The key difference among
these methods is that those adopted in image processing do not use channels to
define watershed lines. Instead, they simply search the landscape for the
highest lines that divide it. In this context, even the most modern
computational models are not able to get all watersheds of the globe at once,
they are usually limited to specific regions. As compared to all these methods,
methods based on invasion percolation as ours do not need to explore the entire
space but only a fractal subset of it. Therefore, our method becomes more and
more efficient the larger the landscape.

Here, we propose a simple model to fully delineate multiple drainage basins for
any given landscape based on the traditional Invasion Percolation (IP) model
\cite{wilkinson1983}, which is known to be a Self-Organized Criticality (SOC)
model \cite{stauffer1994}, {\it i.e.} the IP model does not need a tuning
parameter to evolve and identify the watershed. Our method allows us to study
the morphological segmentation of global maxima and minima in image processing,
even if the landscape stands for the gray scale of a brain Magnetic Resonance
Imaging (MRI) or the heights of a celestial body with or without river channels.
The novelty of our approach is to characterize all basins from a single height
dataset through the definition of a reference (sea) level, {\it i.e.} our
approach is free of parameter tuning.

\section{The Model}

In 2009, Fehr {\it et al.} \cite{fehr2009,fehr2011a,fehr2011b} introduced a
model, called Invasion Percolation-Based Algorithm (IPBA), in order to extract
watersheds from landscapes. The IPBA was proposed for a regular square lattice
of size $L$ with fixed boundary conditions in the vertical direction and
periodic boundary conditions in the horizontal direction, where the height of
each site $i$ was represented by $h_i$. It was also defined that the upper and
lower lines of the lattice represent the sinks of two basins, {\it e.g.} one at
the North (N) and other at the South (S), respectively. In this context, the
following rule for the identification of the basins was proposed: For each site
$i$, one applies the IP model, defining that the basin (N or S) to which the
site $i$ belongs is the one that the IP invaded cluster reaches first. Thus, all
sites of the lattice belong to one of the two basins and the interface line
between them defines the watershed. To improve the computational performance of
the task of finding the interface line, an efficient sweeping strategy was also
introduced: ({\it i}) Initially, the sites are chosen along a straight line that
connects the sinks. Therefore, when the IP processes from two neighbouring sites
evolve to different sinks, a segment of the watershed lies between them. ({\it
ii}) From then on, the sites are chosen only in the neighbourhood of the already
known watershed segments in order to reveal more segments of the watershed,
resulting at the end in the complete watershed. Fehr {\it et al.}
\cite{fehr2009} also showed that the IPBA follows the same dynamics as the
Vincent-Soille algorithm \cite{vincent1991}. The Vincent-Soille algorithm has a
direct interpretation, which is basically a flooding process, although it is
computationally inefficient. The main advantage of the IPBA is its computational
performance, the IPBA presents a sublinear complexity time since it only
explores a fractal subset of space of fractal dimension $91/48$ \cite{fehr2009}
and this characteristic allows us to perform a global analysis to the drainage
basins.

\begin{figure}
\centering
\includegraphics[width=1.0\linewidth]{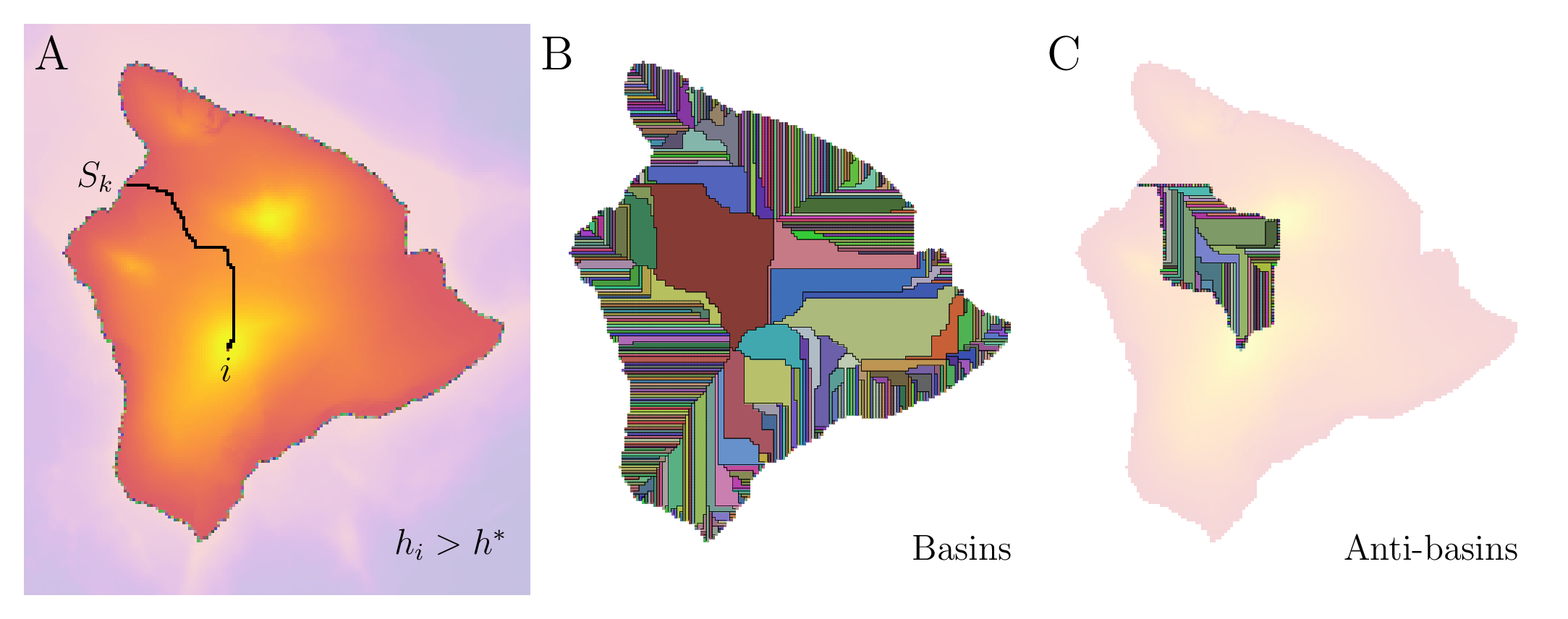}
\caption{Extension of the Invasion Percolation-Based Algorithm (IPBA). We use a
Digital Elevation Model (DEM) from the Island of Hawaii's region to show the
three steps of our approach. (A) Single height cluster composing by all sites
with $h_i>h^*$, where $h^*=0$ (sea level) throughout the study. The landscape
colour is arranged from dark purple (small heights) to light yellow (large
heights) in linear scale. We represent in different colours all sinks, namely,
the sites that are in the height cluster border, in this case, the sea shore.
Finally, the Invasion Percolation (IP) cluster of site $i$ is shown in black.
The IP process starts at the site $i$ and finishes at the sink $S_k$. (B) All
drainage basins identified with our algorithm. The basins have the colours of
their sinks and the black lines stand for the watershed network. (C) All
drainage anti-basins from the largest drainage basin. The anti-basins are also
represented by several colours and the black lines, in this case, stand for the
anti-watershed network.}
\label{fig:ipba}
\end{figure}

Our aim is to define a robust mathematical model for the delineation of multiple
drainage basins through an extension of the IPBA. Suppose a regular rectangular
lattice $L_x \times L_y$, where the height of each site $i$ is $h_i$, analogous
to the original model. We introduce a height threshold $h^*$ such that, if $h_i
> h^*$, then the $i$th site belongs to a cluster, which we call {\it height
cluster}, composed by all connected sites with height above that threshold.
Otherwise, the $i$th site does not belong to any cluster. As explained in the
following section, we adopted $h^*=0$ throughout this study, which for Earth
corresponds to sea level. For this particular choice, the height clusters define
continents and islands on Earth, as shown in Fig.~\ref{fig:ipba}A. Here, the
sinks $S_k$ ($k = 1,2,\dots,N_b$) are all the $N_b$ border sites of the height
clusters, {\it i.e.} the sea shore on Earth. Consequently, we know {\it a
priori} that they define $N_b$ drainage basins separated by several interface
lines, but their specific sizes and shapes need to be determined. Similarly to
the ideas proposed by Fehr {\it et al.} \cite{fehr2009}, we define the following
rule to identify basins present in the height clusters: The IP model is applied
for each site $i$ defining that the basin ($S_k$) at which the site $i$ belongs
is the one that the IP invaded cluster reaches first (see Fig.~\ref{fig:ipba}A).
As depicted in Fig.~\ref{fig:ipba}B, the set of interface lines forms the {\it
watershed network} that separates all basins in the height clusters. We also use
a strategy analogous to the original IPBA to improve the performance of finding
the watershed network. Here, the sweeping occurs in each basin $S_k$ as follows:
({\it i}) The sink $S_k$ defines the ends (initial and final segments) of its
yet not identified watershed. ({\it ii}) For each basin, the sweeping occurs
only at sites neighboring the already known watershed segments in order to
reveal the missing ones. In other words, we scanned the sites along the
watershed inner perimeter neighbourhood of each basin. This strategy drastically
reduces the number of times that we need to apply the IP algorithm for the
identification of the watershed network. ({\it iii}) Optionally, a simple
burning algorithm can be applied to each basin in order to evaluate its area
\cite{stauffer1994}.

Actually, we consider two versions of our algorithm along the study: A version
with traditional periodic boundary conditions in horizontal direction and
unconventional periodic boundary conditions in vertical direction for real
landscapes, and another version with fixed boundary conditions on both
directions for artificial landscapes. The unconventional periodic boundary
conditions, adopted for real landscapes, are defined by imposing that each site
in the top (bottom) row be neighbour of every other site in the top (bottom)
row. These boundary conditions represent a mapping of a sphere into a lattice.
They are needed because we performed all measures on the entire globe for real
landscapes.

A natural extension of the watershed concept is the definition of its reciprocal
line, called anti-watershed. The {\it anti-watersheds} are composed by lines of
minimal heights, in contrast to the watersheds, defined in terms of lines of
maximal heights. Therefore, if the basins can (grossly) be understood as
``cavities'' in a surface, then the anti-basins can also be understood as
``humps'' on that same (inverted) surface. We emphasize that anti-watersheds do
not necessarily represent river channels, since a river channel (or a channel)
is a ``clearly defined watercourse (``natural or man-made channel through or
along which water may flow'') which periodically or continuously contains moving
water'', or is a ``watercourse forming a connecting link between two water
bodies'', or even is the ``deepest portion of a watercourse, in which the main
stream flows'' \cite{wmo2012}, while a anti-watershed line may never contain any
flowing water. Given an upside-down landscape, the same approach for watersheds
can be considered to define the anti-watersheds. In this case, the watershed
network represents the {\it anti-watershed network} in the original landscape.
Furthermore, we can also define an anti-watershed network within a drainage
basin (see Fig.~\ref{fig:ipba}C). Here, the lines of minimal heights represent
the most deeper rivers and their tributaries in a lot of situations.

We highlight that our model just takes into account the drainage basins that
flow out to the oceans, {\it i.e.} we are ignoring the {\it endorheic drainage
basins} (or simply {\it endorheic basins}), which are those basins that flow out
to any place other than the oceans, {\it e.g.} lakes or swamps, their amounts of
water being balanced by infiltration or evaporation \cite{busby2012}.
Nonetheless, we could also include such basins in our model introducing
additional sinks $S_k'$ (where $k =1,2,\dots,N_e$ and $N_e$ is the number of
endorheic basins), located at the points of higher evaporation rates, for
example. In this context, the total number of drainage basins ($N_b+N_e$) would
depend on the spatial resolution of the data and the information available about
the endorheic basins. Our code is available at \url{https://github.com/erneson}.

\section{Results}

We applied the IPBA model to real and artificial landscapes in order to study
the statistical properties of the drainage basins on Earth, Moon, and Mars. For
real landscapes, we use three different DEMs throughout this study: the General
Bathymetric Chart of the Oceans (GEBCO) \cite{gebco2014}, the Lunar Orbiter
Laser Altimeter (LOLA) \cite{lola2014}, and the Mars Orbiter Laser Altimeter
(MOLA) \cite{mola2014}. Such datasets consist of the map of heights for the
Earth, Moon and Mars, respectively. Moreover, we obtained the artificial
landscapes through the {\it fractional Brownian motion} (fBm) \cite{fisher1988}.
We chose this method because we are interested in generating landscapes with
tunable spatial long-range correlations since it is known that such correlations
change the statistical properties of watersheds \cite{fehr2011b}. We show all
landscapes in Figs.~\ref{fig:gebco_lola_mola}A-C and Figs.~\ref{fig:fbm2d}A-D.
We considered two scenarios for GEBCO, LOLA, and MOLA datasets: the original and
the upside-down landscape orientations. We emphasize that we chose the GEBCO
dataset to represent the terrestrial landscapes because it includes altimetric
and bathymetric heights, which makes it more similar to LOLA and MOLA datasets.
In other words, we used the GEBCO dataset in order to make a general and uniform
comparison between Earth and two celestial bodies that do not have oceans (Moon
and Mars). To perform a global analysis, the resolutions of the real datasets
were decreased by a factor of $r=8$, where $r$ is the resolution factor. Thus,
each tile of $8 \times 8$ sites was replaced by a single site with a value given
by the mean of all $64$ original sites. The sizes of the used lattices were
$5400 \times 2700$ for GEBCO (original $43200 \times 21600$), $5760 \times 2880$
for LOLA (original $46080 \times 23040$), and $5760 \times 2880$ for MOLA
(original $46080 \times 23040$). For the fBm landscapes, we considered only the
original orientation due to the natural symmetry of the Gaussian distribution
used in the Ffm. In this case, we averaged our simulations over $10$ samples of
lattices of $4096 \times 4096$ for the traditional range of the Hurst exponent
$H$ ($0\le H \le 1$). The following subsections show the corresponding results.

\subsection{Real Landscapes}

\begin{figure}
\centering
\includegraphics*[width=1.0\textwidth]{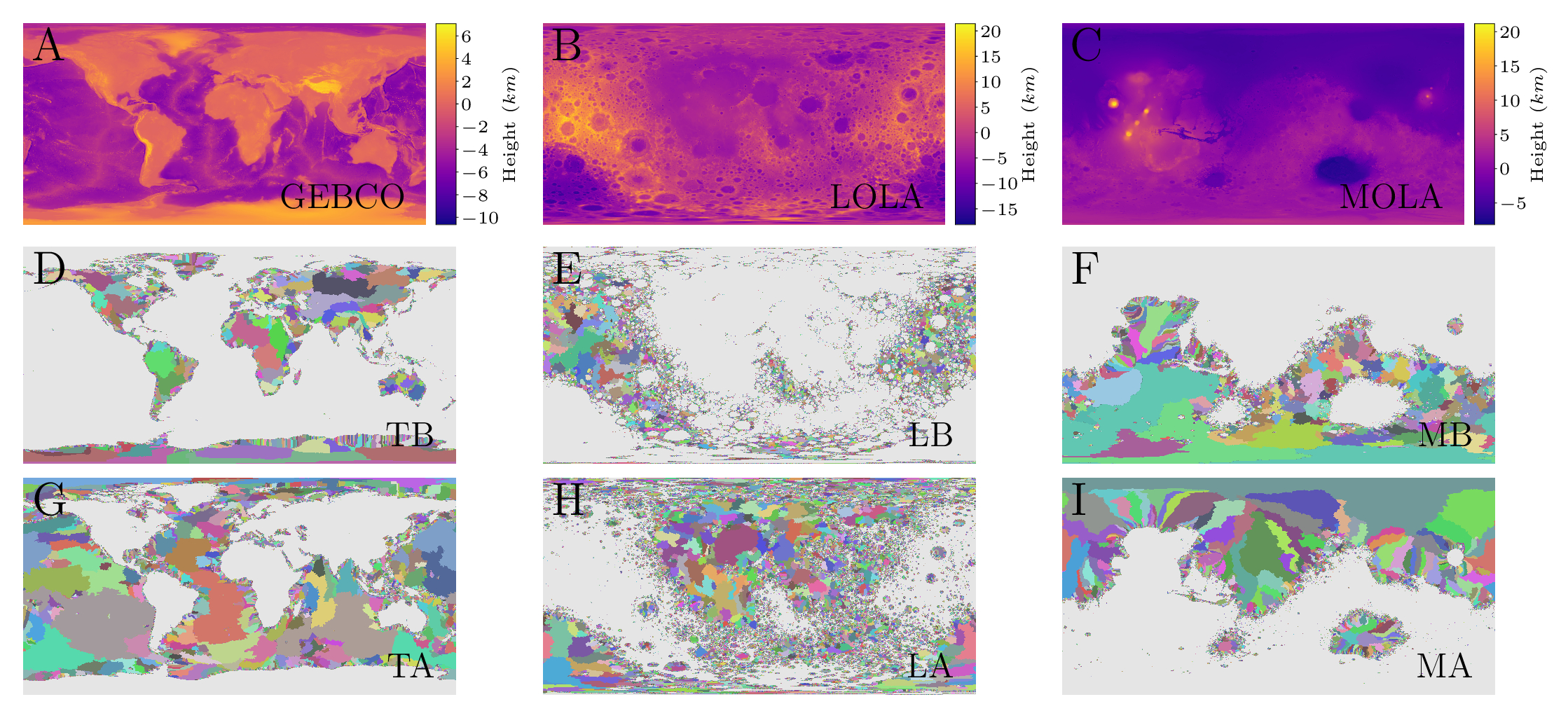}
\caption{Landscape, basins and anti-basins for Earth, Moon, and Mars. (A)
General Bathymetric Chart of the Oceans (GEBCO) \cite{gebco2014}. (B) Lunar
Orbiter Laser Altimeter (LOLA) \cite{lola2014}. (C) Mars Orbiter Laser Altimeter
(MOLA) \cite{mola2014}. In all three cases, the heights are in units of
kilometre ($km$) and represented in linear scale. (D-F) Drainage basins for
heights above $h^*$ of the GEBCO, LOLA, and MOLA landscapes. (G-I) Drainage
anti-basins for heights below $h^*$ of the GEBCO, LOLA, and MOLA landscapes.
Note that $h^*=0$ (sea level for Earth and hypothetical sea level for Moon and
Mars) throughout the study. Here, we use the following abbreviations:
Terrestrial Basins (TB), Terrestrial Anti-basins (TA), Lunar Basins (LB), Lunar
Anti-basins (LA), Martian Basins (MB), and Martian Anti-basins (MA). Sites below
the height threshold $h^*$ are shown in grey.}
\label{fig:gebco_lola_mola}
\end{figure}

Here, we show the main results of our approach applied to real landscapes. In
Figs.~\ref{fig:gebco_lola_mola}D and G, we show the results of our model applied
to the original and upside-down landscapes of the Earth. In
Figs.~\ref{fig:gebco_lola_mola}E, H, F, and I, we use the height threshold
$h^*=0$ (hypothetical sea level) and perform the same analysis used for Earth's
landscape to obtain the basins and anti-basins for the Moon and Mars. We note
that the basins and anti-basins look similar in the terrestrial and martian
landscapes, while in the lunar landscape, they are affected by the {\it impact
basins}, {\it i.e.} craters originated from the impact of asteroids. This
similarity is quantified here through the statistical distributions of
perimeters and areas of the basins and anti-basins for Earth, Moon and Mars. We
included the lunar analysis as a counterpoint to show that the observed
similarity between the terrestrial and Martian results is indeed genuine. The
log-log plots of all these distributions shown in Figs.~\ref{fig:g_p_A}A-B
clearly indicate the presence of long tails extending over several orders of
magnitude. Moreover, these tails approximately follow power-law behaviors, $P(s)
\propto s^{\alpha}$ and $P(A) \propto A^{\beta}$, for the perimeters and areas,
respectively. By performing Ordinary Least Square (OLS) fits
\cite{montgomery2006} to the corresponding distributions, we obtained estimates
for the power-law exponents $\alpha$ and $\beta$ that are summarized in
Table~\ref{tab:t_exp}. Similar results where found using a Maximum Likelihood
Estimator (MLE) \cite{clauset2009}, as shown in the Supplementary Information
(SI).

We found that the terrestrial and martian results are statistically identical,
which suggests the surfaces of both planets underwent a similar formation
history and that a hypothetical martian river would present some level of
similarity to the terrestrial rivers, since both landscapes share the same
statistics for watershed (maxima) and anti-watershed (minima) lines. There is
already evidence that the Martian channels were formed by a series of fluvial
and non-fluvial processes \cite{hargitai2018}. Furthermore, we found evidence
that the obtained perimeter and area exponents are independent of the resolution
factor $r$ (See SI). It is known that the effects of finite size are attenuated
as the system size increases. We also obtained similar results for other values
of $h^*$ on Earth, Moon and Mars.

\begin{table}
\centering
\begin{tabular}{c|c|c|c}
Cases                   & Abbreviation & $\alpha$        & $\beta$\\
\hline
Terrestrial Basins      & TB          & 2.30 $\pm$ 0.03 & 1.74 $\pm$ 0.02\\
Terrestrial Anti-basins & TA          & 2.26 $\pm$ 0.03 & 1.72 $\pm$ 0.02\\
Lunar Basins            & LB          & 2.70 $\pm$ 0.02 & 1.93 $\pm$ 0.01\\
Lunar Anti-basins       & LA          & 2.64 $\pm$ 0.04 & 1.89 $\pm$ 0.02\\
Martian Basins          & MB          & 2.33 $\pm$ 0.04 & 1.75 $\pm$ 0.02\\
Martian Anti-basins     & MA          & 2.26 $\pm$ 0.02 & 1.73 $\pm$ 0.02\\
\end{tabular}
\caption{Exponents $\boldsymbol{\alpha}$ and $\boldsymbol{\beta}$ for real
landscapes. These exponents were determined through the Ordinary Least Square
(OLS) fits \cite{montgomery2006} to the corresponding distributions. Similar
results were found using a Maximum Likelihood Estimator (MLE) \cite{clauset2009}
(See SI).}
\label{tab:t_exp}
\end{table}

\begin{figure}
\includegraphics*[width=1.0\linewidth]{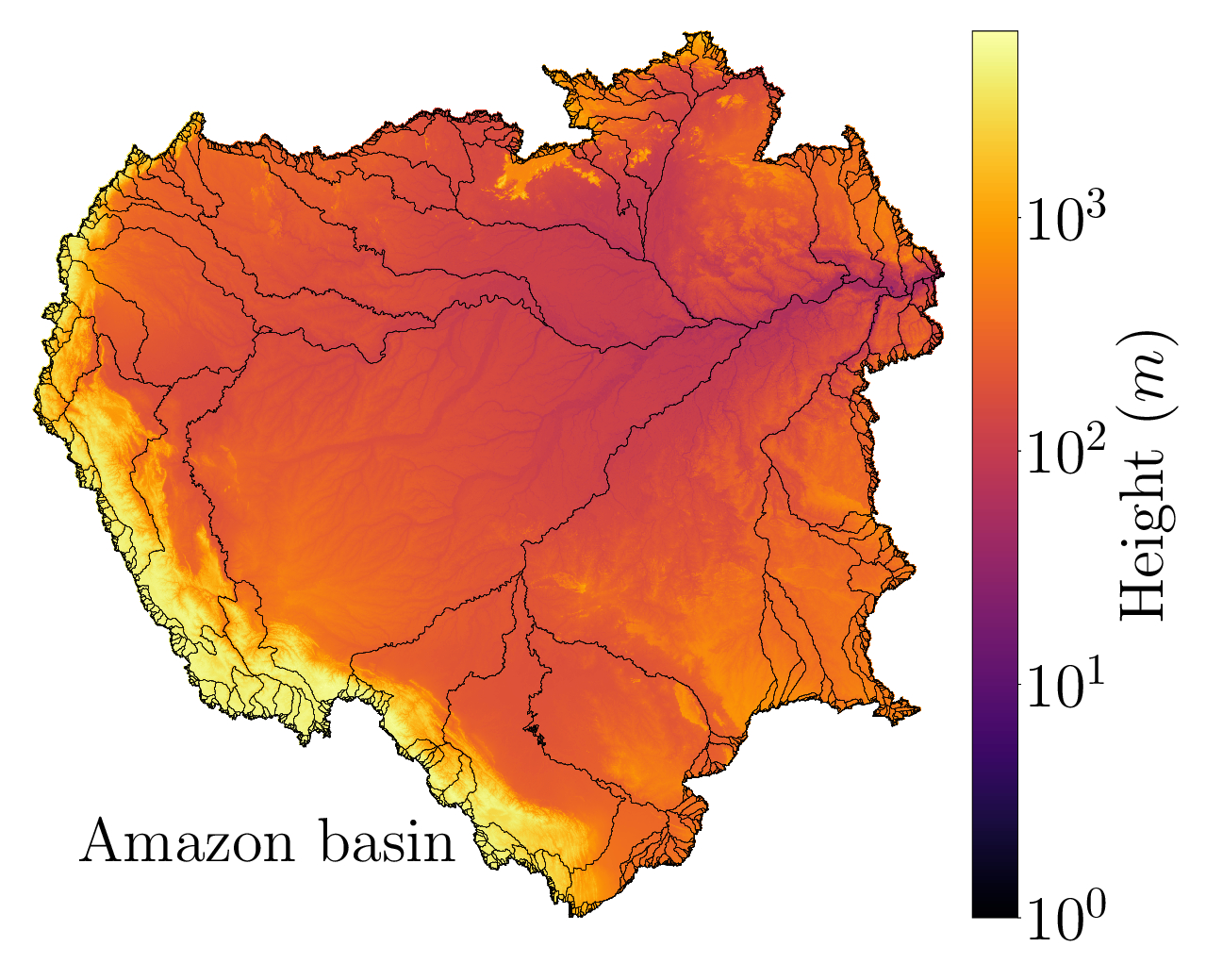}
\caption{The Amazon basin. The drainage basin of the Amazon region obtained by
our algorithm. The heights are represented in metres ($m$) and in logarithmic
scale in order to show the details of the Andes mountain range and the Amazon
river. The black lines stand for the anti-watershed network, which shows an
impressive similarity with the Amazon river network.}
\label{fig:amazon}
\end{figure}

In Fig.~\ref{fig:amazon}, we show the Amazon basin and its associated
anti-basins defined by our algorithm on GEBCO dataset at the original
resolution. In this special case, we are removing all internal sinks from the
South American continent, {\it i.e.} we are allowing the existence of sites with
negative heights within South America. We emphasise that several rivers (the
deeper ones) follow the anti-watershed lines. The black lines in
Fig.~\ref{fig:amazon} stand for the anti-watershed network, which shows an
impressive similarity with the Amazon river network (See SI for further
comparisons). This similarity allows us to obtain the length of the longest
river in a basin by approximating it by the length of the longest path from the
point of minimal height (the mouth of the river) of the largest tree on the
anti-watershed network. We defined the height of each point of the
anti-watershed lines as the mean of the heights of its neighbouring sites.
Therefore, we are ignoring any issues related to the initiation point of river
channels \cite{wohl2017}. As a perspective for future work, quantitative
analyses can be performed comparing the anti-watershed and river networks, {\it
e.g.} see the proposed method by Grieve {\it et al.} \cite{grieve2016}.

We also performed the verification of Hack's law \cite{rodriguez-iturbe2001} for
the entire planet. This scaling law establishes the relation between the areas
($A$) of the basins and the maximum lengths ($\ell$) of their rivers, {\it i.e.}
\begin{equation}\label{hack}
\ell \propto A^{\gamma},
\end{equation}

\noindent
where $\gamma$ is known as the Hack exponent. In Figure~\ref{fig:g_ell_A}, we
show Hack's law for the original orientation of the terrestrial landscape
considering only the basins with area greater than $A^*=100\;km^2$. We chose an
area threshold $A^*$ to ensure that the analyzed basins were not too much
affected by the resolution of the dataset. We found the Hack exponent
$\gamma=0.54 \pm 0.01$ with coefficient of determination $R^2=0.87$, which is
very close to our theoretical value of $\gamma \approx 0.55$ for Earth.

\subsection{Artificial Landscapes}

\begin{figure}
\centering
\includegraphics*[width=1.0\linewidth]{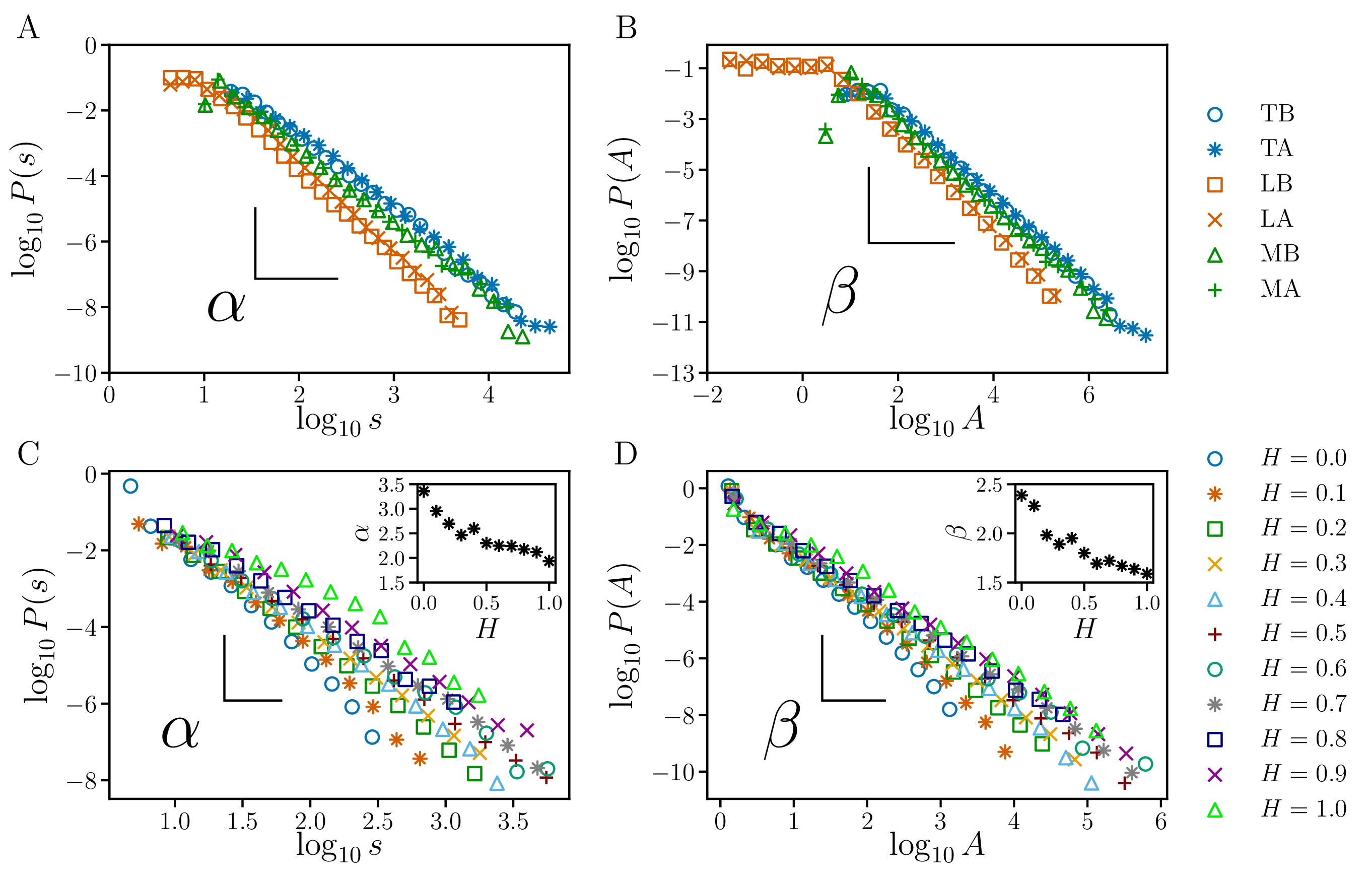}
\caption{Log-log plots of the perimeter and area distributions for basins and
anti-basins on real and artificial landscapes. (A) The perimeter distributions
for basins and anti-basins on Earth, Moon and Mars. (B) The surface area
distributions for basins and anti-basins on Earth, Moon and Mars. In both, we
use the following abbreviations: Terrestrial Basins (TB), Terrestrial
Anti-basins (TA), Lunar Basins (LB), Lunar Anti-basins (LA), Martian Basins
(MB), and Martian Anti-basins (MA). We show all exponents for real landscapes in
Table~\ref{tab:t_exp}. (C) The perimeter distributions for several values of the
Hurst exponent $H$. (D) The surface area distributions for several values of
$H$. The insets show the behaviour of the exponents $\alpha$ and $\beta$ in
relation to $H$. All exponents are calculated through the Ordinary Least Square
(OLS) fits \cite{montgomery2006}. We also obtain similar exponents using a
Maximum Likelihood Estimator (MLE) \cite{clauset2009} (See SI).}
\label{fig:g_p_A}
\end{figure}

We applied the extension of the IPBA to artificial landscapes. In
Figs.~\ref{fig:fbm2d}E-H, we show all basins defined by our model for the same
samples presented in Figures~\ref{fig:fbm2d}A-D. As shown in
Figures~\ref{fig:g_p_A}C and D, the perimeter and area distributions obtained
from these landscapes are systematically affected by the presence of spatial
correlations, quantified here in terms of the parameter $H$. Moreover, except
for the case of $H=-1$, all distributions generated from artificial landscapes
can be approximately described by power laws, $P(s) \propto s^{-\alpha}$ for
perimeters, and $P(A) \propto A^{-\beta}$ for areas. For each value of $H$, we
averaged both distributions for all samples. The insets of the Figures
\ref{fig:g_p_A}C and D show that $\alpha$ and $\beta$ decrease with the Hurst
exponent $H$. The exponents $\alpha$ and $\beta$ range between $3.36$ and $1.93$
and between $2.39$ and $1.59$, respectively. In the uncorrelated case ($H=-1$),
however, we obtained less than one order of magnitude for $\ell$ and $A$
precluding the same kind of analysis.

\begin{figure}
\centering
\includegraphics*[width=0.8\linewidth]{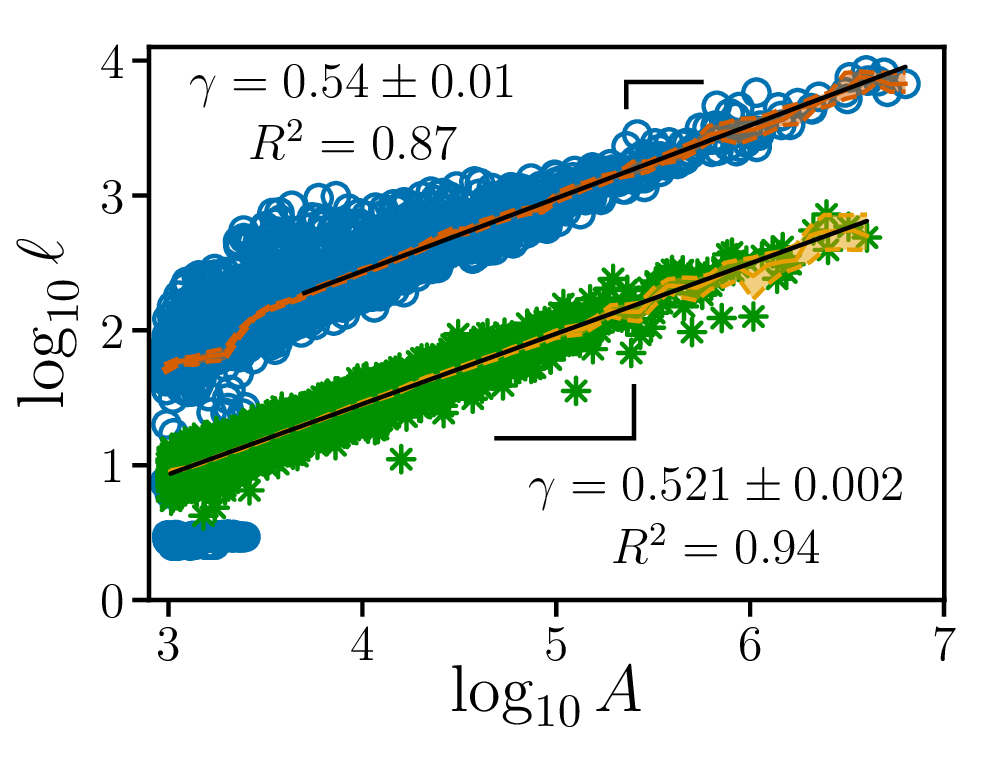}
\caption{Hack's law for Earth and for fBm landscapes with $H=0.7$. Scaling of
the longest path $\ell$ from the point of minimal height (mouth) of the largest
tree on the anti-watershed network versus the area $A$ of its basin (blue
circles for Earth and gree asteriks for fBm landscapes), in square kilometres
($km^2$) for Earth. The solid orange and yellow lines are the Nadaraya-Watson
estimator \cite{nadaraya1964, watson1964}, the orange and yellow shaded regions
are bounded by the lower and upper confidence intervals, and the solid black
line is the linear regression calculated via Ordinary Least Square (OLS)
\cite{montgomery2006}. For Earth, the OLS was applied only to
basins greater than $5,000\;km^2$.}
\label{fig:g_ell_A}
\end{figure}

In Fig. \ref{fig:g_ell_A}, we show Hack's law obtained for $H=0.7$ (a close
value of $H$ is usually obtained for real landscapes \cite{fehr2011b}),
considering only the basins with area above $A^*=1024$. Such result led us to
the following conjecture: Let the basin area be $A$, the longest anti-watershed
line be $\ell$, and assuming that the anti-watershed lines are indeed watershed
lines of the upside-down landscapes, it is known that $\ell \propto L^D$, where
$L$ is the linear length of the system and $D$ is the fractal dimension of the
watershed lines \cite{fehr2009}. Since $A \propto L^2$, we have:

\begin{equation}\label{hack_finite_size}
L^D \propto \ell \propto A^{\gamma} \propto L^{2\gamma},
\end{equation}

\noindent
which gives $\gamma=D/2$. In other words, the Hack exponent depends on the
fractal dimension of the anti-watershed lines. Fehr {\it et al.}
\cite{fehr2011b} showed that the fractal dimension of the watersheds decreases
with the Hurst exponent $H$, similarly to the Optimal Path Cracks
\cite{oliveira2011}, and the coastlines on correlated landscapes
\cite{morais2011}. The fractal dimension of the watershed lines ranges between
$D=1.0$ and $D=1.22$, what gives to the Hack exponent a corresponding range from
$\gamma=0.5$ to $\gamma=0.61$. For real landscapes, the fractal dimensions of
the watershed lines are around $1.10$ ($D=1.10$ for the Alps \cite{fehr2009},
$D=1.11$ for the Himalaya \cite{fehr2009}, and $D=1.12$ for the Andes
\cite{fehr2011b}). Therefore, our expected Hack exponent for Earth should be
$\gamma \approx 0.55$, a value which is in good agreement with the result shown
in Fig. \ref{fig:g_ell_A} ($\gamma=0.521 \pm 0.002$ with coefficient of
determination $R^2=0.94$ for artificial landscapes). This result suggests that
Hack's law, often observed for river networks, is an intrinsic effect of
topography, {\it i.e.} it depends, in essence, on the watershed and
anti-watershed lines. In other words, Hack's law may have a purely geometrical
origin and does not depend on physical laws governing the water flow on a
surface \cite{birnir2008}.

\section{Discussion}

We proposed a general model to fully delineate multiple drainage basins for any
given landscape of heights through an extension of the IPBA. The novelty of our
approach is to characterise all basins from a single height dataset through the
definition of a reference (sea) level. Such fact allows us to claim that our
model is free of parameter tuning. In this way, we are able to delineate the
basins through the definition of the watershed network (maximal lines of a
landscape) as well as the anti-basins through the definition of the
anti-watershed network (minimal lines of a landscape). In order to show that our
algorithm was robust, we applied it to real and artificial landscapes. In both
cases, we found that the perimeter and area distributions are ruled by power
laws with exponents $\alpha$ and $\beta$, respectively. It was also shown that
the terrestrial and martian results are statistically identical, which suggests
that the surfaces of Earth and Moon have undergone similar formation processes
and that a hypothetical martian river would present similarity to the
terrestrial rivers, since both landscapes share the same statistics for
watershed and anti-watershed networks. We also verified that, in the Amazon
basin and its associated anti-basins defined by our approach, several rivers
(the most deeper ones) rest on anti-watershed lines. Furthermore, we showed that
the exponents $\alpha$ and $\beta$, for artificial landscapes, decrease
systematically with the Hurst exponent $H$ and that they are invariant under the
inversion of real landscapes. Finally, we found a theoretical value for the
Hack's exponent based on the fractal dimension of the watershed and
anti-watershed lines, $\gamma=D/2$. We measured $\gamma=0.521 \pm 0.002$ for
artificial landscapes with $H=0.7$ and $\gamma=0.54 \pm 0.01$ for Earth, which
agree within error bars with our estimation of $\gamma \approx 0.55$ for real
cases.

\section{Methods}

\subsubsection{Real Landscapes}

We use three different DEM datasets throughout this study. The first dataset is
the General Bathymetric Chart of the Oceans (GEBCO) \cite{gebco2014} consisting
of altimetric and baltimetric heights, {\it i.e.} the heights above and below
the sea level, around the Earth globe. The resolution of this dataset is
$30\;\text{arc-seconds}$ ($30/3600\;\text{decimal degrees}$) in both
coordinates, equivalent to a square lattice with edge length of
$0.926\;\text{kilometers}\;(km)$ at the Equator line. The other two are the
Lunar Orbiter Laser Altimeter (LOLA) \cite{lola2014} and Mars Orbiter Laser
Altimeter (MOLA) \cite{mola2014} consisting of the map of heights for the Moon
and Mars, respectively. The LOLA resolution is about $0.118\;km$, while MOLA
resolution is $0.463\;km$, both in relation to their corresponding ``Equator''.
We show the three datasets in Figs.~\ref{fig:gebco_lola_mola}A-C. In addition,
we point out that all datasets are freely available and are friendly
ready-to-use, {\it i.e.} all technical preprocessing steps were already
performed by the GEBCO, LOLA, and MOLA research teams. We do not perform any
additional preprocessing (such as any hydrologic correction or void filling),
except the addition of a tiny noise in the heights ($<0.001\;m$, much less than
the precision of all datasets) in order to ensure that all values in real
landscapes are different, defining a unique sequence if the height values are
sorted.

We also emphasise that all three datasets are available in the Geographic
Coordinate System (GCS), more precisely, in latitude and longitude grids in the
image format {\it TIFF}, {\it i.e.} they are mapped on spheres, the terrestrial
(with radius $R_{earth}=6378.137\;km$), the lunar (with radius
$R_{moon}=1737.4\;km$), and the martian (with radius $R_{mars}=3396.19\;km$).
For GEBCO, the reference surface (the zero height) is defined by the terrestrial
geoid. The geoid is the natural shape that a static fluid would present due to
the gravitational potential of its celestial body \cite{schubert2007}. On Earth,
the oceans could be considered static and, consequently, they are well
approximated by such a surface. For LOLA and MOLA, the geoid concept is
generalised by the gravitational equipotential surface with the mean lunar and
martian radius at the Equator, respectively, defining hypothetical sea levels
\cite{lola2014,mola2014}. We adopt the height threshold $h^*=0$ for all
landscapes in order to make a general comparative analysis.

Here, we perform the calculation of the area of each site by the composition of
two spherical triangles (the site areas for artificial landscapes have no unit
of measure and are all unitary). The area of a spherical triangle with edges
$a$, $b$ and $c$ is given by \cite{todhunter1863},
\begin{equation}\label{A}
A=4 R_k^2 \tan^{-1}\left[\tan\left(\frac{s}{2}\right)\tan\left(\frac{s_a}{2}\right)\tan\left(\frac{s_b}{2}\right)\tan\left(\frac{s_c}{2}\right)\right]^{1/2},
\end{equation}

\noindent
where $s=(a/R_k+b/R_k+c/R_k)/2$, $s_a=s-a/R_k$, $s_b=s-b/R_k$, and
$s_c=s-c/R_k$. In this formalism, $R_k$ is the sphere radius, where, in our
case, $k=\{earth,moon,mars\}$, and the edge lengths are calculated by the great
circle (geodesic) distance between two points $i$ and $j$ on the sphere surface
given by the {\it Haversine formula} \cite{snyder1987}:

\begin{equation}\label{dij}
d_{ij}=2R_k\sin^{-1}\left[\sqrt{\sin^2\left( \frac{\Delta\phi}{2} \right)+\cos\phi_i \cos\phi_j\sin^2\left( \frac{\Delta\lambda}{2} \right)}\right],
\end{equation}

\noindent
where $\Delta\phi=\phi_j-\phi_i$ and $\Delta\lambda=\lambda_j-\lambda_i$. The
values of $\lambda_i$ ($\lambda_j$) and $\phi_i$ ($\phi_j)$, measured in
radians, are the longitude and latitude, respectively, of the point $i$ ($j$).
Therefore, we are able to define the site areas, and, consequently, obtain the
total area of a given basin, since each basin is composed by a set of sites.

\subsubsection{Artificial Landscapes}

\begin{figure}
\centering
\includegraphics*[width=1.0\linewidth]{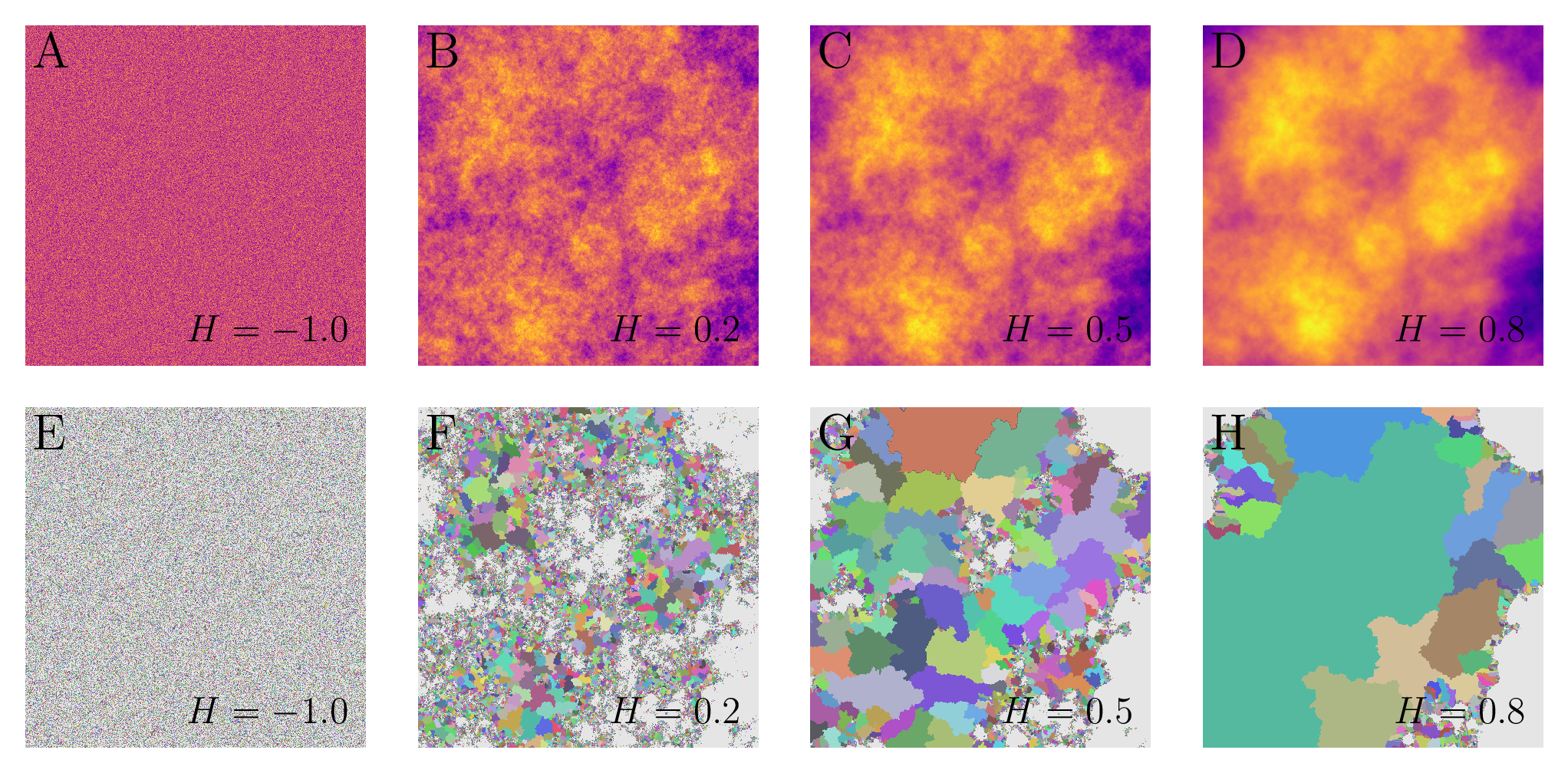}
\caption{Artificial landscapes and their respective basins. Fractional Brownian
motion (fBm) landscapes generated by the Fourier filtering method (Ffm) for four
regimes. A typical uncorrelated landscape ($H=-1$) is shown in (A), a landscape
generated with negative correlations ($H=0.2$) in (B), a Brownian motion
landscape ($H=0.5$) in (C), and a landscape with positive correlations ($H=0.8$)
in (D). All fBm landscapes shown share the same size $L=1024$ and the same
random seed. The heights are represented in linear scale, where lighter (darker)
colours stand for higher (lower) heights. The basins defined by our approach
corresponding to the fBm landscapes shown in (A), (B), (C) and (D) are shown in
(E), (F), (G) and (H), respectively. Sites below the height threshold $h^*$ are
show in grey.}
\label{fig:fbm2d}
\end{figure}

We obtained artificial landscapes through the {\it fractional Brownian motion}
(fBm) \cite{fisher1988} in order to study the watershed and anti-watershed
networks. One of the most established method to generate a fBm is the so-called
{\it Fourier filtering method} (Ffm) \cite{fisher1988}. The basic idea of the
Ffm is to define random Fourier coefficients in the reciprocal space,
distributed according to the following power-law spectral density:
\begin{equation}\label{spectraldensity}
    S(f_1,f_2,\dots,f_d) = \sqrt{\left(\sum_{i=1}^d f_i^2\right)^{-w}},
\end{equation}

\noindent
where $f_i$ is the frequency of the dimension $i$, $d$ is the topological
dimension, and $w$ is the spectral exponent. Subsequently, the inverse Fourier
transform is applied to generate a correlated distribution in the real space. In
our case $d=2$, the correlated distribution is a landscape. Each landscape is
characterised by an exponent $H$, called {\it Hurst exponent}, related to the
spectral exponent by $w=2H+d=2H+2$. Four cases can be distinguished: ({\it i})
For $H=-1$, the uncorrelated landscape (see Fig.~\ref{fig:fbm2d}A). ({\it ii})
For $0<H<1/2$, the landscape has a negative correlation, {\it i.e.} the
increments are anticorrelated (see Fig.~\ref{fig:fbm2d}B). ({\it iii}) For
$H=1/2$, the landscape is correlated, but the increments are uncorrelated (see
Fig.~\ref{fig:fbm2d}C), which is the case of the classical {\it Brownian motion}
\cite{fisher1988}. ({\it iv}) Finally, for $1/2<H<1$, the landscape has a
positive correlation, {\it i.e.} the increments are correlated (see
Fig.~\ref{fig:fbm2d}D).

\section{Acknowledgements}

We gratefully acknowledge CNPq, CAPES, FUNCAP and the National Institute of
Science and Technology for Complex Systems in Brazil for financial support.

\section{Author contributions statement}

E.A.O., H.J.H., and J.S.A. designed research; E.A.O., R.S.P., and R.S.O.
performed research; E.A.O., R.S.P. and J.S.A. analyzed data; and E.A.O., R.S.P.,
R.S.O., V.F., H.J.H., and J.S.A. wrote the paper. All authors reviewed the
manuscript.

\section{Data Availability}

All data used in this manuscript are free available. Please, check the
references related to GEBCO, LOLA, and MOLA datasets.

\section{Additional information}

\subsection{Competing interests}

We declare we have no competing interests.

\clearpage

\end{document}